 \documentclass[structabstract]{aa}
  \usepackage{txfonts}
   \usepackage{graphicx}
    \usepackage{natbib}
     \usepackage{longtable}
      \usepackage{subeqnarray}
       \usepackage{cases}
\begin{document}

\title {The relation of H$_2$CO, $^{12}$CO, and $^{13}$CO in molecular clouds\thanks{Appendices are only available in electronic form at http://www.aanda.org.}}

\author{Xin. Di. Tang\inst{1,2}
 \and Jarken. Esimbek\inst{1,3}
  \and Jian. Jun. Zhou\inst{1,3}
   \and Gang. Wu\inst{1,3}
    \and Wei. Guang. Ji\inst{1,2}
     \and Daniel. Okoh\inst{1,4}}


\institute{Xinjiang Astronomical Observatory, Chinese Academy of Sciences, Urumqi 830011, PR China\\
 \email{tangxindi@xao.ac.cn}
  \and Graduate University of the Chinese Academy of Sciences, Beijing 100080, PR China\\
   \and Key Laboratory of Radio Astronomy, Chinese Academy of Sciences, Urumqi 830011, PR China \\
    \and Physics \& Astronomy Department, University of Nigeria, Nsukka 410001, Nigeria}

\date{Received / Accepted  }

\abstract
{}
{We seek to understand how the 4.8 GHz formaldehyde absorption line is distributed in the MON R2, S156, DR17/L906, and M17/M18 regions. More specifically, we look for the relationship among the H$_2$CO, $^{12}$CO, and $^{13}$CO spectral lines.}
{The four regions of MON R2 (60\arcmin{} $\times$ 90\arcmin), S156 (50\arcmin{} $\times$ 70\arcmin), DR17/L906 (40\arcmin{} $\times$ 60\arcmin), and M17/M18 (70\arcmin{} $\times$ 80\arcmin) were observed for H$_2$CO (beam 10\arcmin), H110$\alpha$ recombination (beam 10\arcmin), 6 cm continuum (beam 10\arcmin), $^{12}$CO (beam 1\arcmin), and $^{13}$CO (beam 1\arcmin). We compared the H$_2$CO, $^{12}$CO, $^{13}$CO, and continuum distributions, and also the spectra line parameters of H$_2$CO, $^{12}$CO, and $^{13}$CO. Column densities of H$_2$CO, $^{13}$CO, and H$_2$ were also estimated.}
{We found out that the H$_2$CO distribution is similar to the $^{12}$CO and the $^{13}$CO distributions on a large scale. The correlation between the $^{13}$CO and the H$_2$CO distributions is better than between the $^{12}$CO and  H$_2$CO distributions. The H$_2$CO and the $^{13}$CO tracers systematically provide consistent views of the dense regions. Their maps have similar shapes, sizes, peak positions, and molecular spectra and present similar central velocities and line widths. Such good agreement indicates that the H$_2$CO and the $^{13}$CO arise from similar regions.}
{}

\keywords{interstellar medium: molecular clouds -- interstellar medium: molecules -- stars: formation}
 \maketitle

\section{INTRODUCTION}

The H$_2$CO distribution in the Galaxy has been noted by Davies \& Few (1979). They found that the H$_2$CO distribution is similar to that of \ion{H}{ii} and CO. Downes et al. (1980) surveyed 262 Galactic radio sources using the H$_2$CO absorption line at 4.830 GHz, and H110$\alpha$ recombination line at 4.874 GHz, and found that H$_2$CO is associated with most of the \ion{H}{ii} regions. It is a good probe of the star formation region. H$_2$CO absorption is only seen in absorption against the background continuum, and it gives different constraints to mm and sub-mm spectral lines, which are seen both in front of and behind the \ion{H}{ii} region. It provides a unique probe of the physical conditions for the foreground cloud. Comparative surveys of H$_2$CO absorption and CO emission in the galactic center have been reported by Scoville et al. (1972, 1973) and Solomon et al. (1972). A good correlation is generally found between H$_2$CO and CO. Cohen et al.(1983) generally found good agreement on a large scale but a connection that does not have enough detail. Recently, Rodr\'{\i}guez et al.(2006, 2007) and Zhang et al. (2012) compared the H$_2$CO absorption and CO emission profiles towards the Galactic anticenter, and the following five regions, L1204/S140, W49, W3, DR21/W75, and NGC2024/NGC2023. They found a crude correlation between these two molecular tracers on a large scale. Generally, the $^{12}$CO (1--0) emission line is optically thick, and the H$_2$CO (1$_{10}$--1$_{11}$) absorption line is optically thin, so the two lines have many different properties in the dense region. The $^{13}$CO (1--0) emission line is optically thin, and it can trace dense region (n(H$_2$) $>$ 10$^{3}$ cm$^{-3}$). It is therefore similar to H$_2$CO.

Liszt \& Lucas (1995) compared N(H$_2$CO) with N(HCO$^+$) and N($^{13}$CO) towards compact extragalactic mm-wave continuum sources, and found that H$_2$CO has a rapid increase with N(HCO$^+$), but not as rapid as that of N($^{13}$CO). N($^{13}$CO) is strongly affected by fractionation (Liszt \& Lucas 1998), so that the H$_2$CO-$^{13}$CO comparison is perhaps interesting, while it is not directly used in deciphering the general chemistry. Cohen et al. (1983) compared the H$_2$CO (beam $\sim$ 10\arcmin) with $^{13}$CO (beam $\sim$ 8\arcmin) maps in the Orion molecular cloud and found that the agreement between H$_2$CO and $^{13}$CO is considerably better, but with different relative intensities and with minor differences in the detailed morphology. Therefore, making a point-by-point comparison will be interesting when the $^{13}$CO profiles become available. In this paper, we report new CO, H$_2$CO, 6 cm continuum and H110$\alpha$ observations in four Galactic HII regions of MON R2, S156, DR17/L906, and M17/M18. We are interested in making a comparative study of the H$_2$CO, $^{12}$CO, and $^{13}$CO lines.

\section{OBSERVATIONS}
\subsection{Formaldehyde, H110$\alpha$ recombination, and continuum}

From September 2010 to August 2011, we observed the H$_2$CO line, the H110$\alpha$ line, and the 6 cm continuum with the Nanshan 25 m radio telescope of Xinjiang Astronomical Observatory. The 25 m radio telescope has an HPBW (half power beam width) of 10\arcmin{} at 4829.6594 MHz. A 6 cm low noise receiver was used. The system temperature was about 23 K during observations. Digital autocorrelation spectrometers were used with 4096 channels and 80 MHz bandwidth, and a corresponding velocity resolution of 1.206 km s$^{-1}$. The line detection limit is about 23 mK with 10 min integration time. To get the higher signal/noise ratio, each point's total integration time ranged from 30 min to several hours. The continuum at 4.8 GHz were processed with a bandwidth of about 400 MHz, and the error was approximately 5\%. The DPFU (degrees per flux unit) value was 0.116 K Jy$^{-1}$. The pointing and tracking accuracy was better than 15\arcsec, and the beam efficiency was 65\%. The observation was performed in the so-called ON/OFF mode.  A diode noise source was used to calibrate the spectrum and the flux error was 15\%.

\subsection{Carbon monoxide}

From 15 to 26 May 2011, the $^{12}$CO and $^{13}$CO observations of the four regions were carried out with the 13.7 m millimeter wave telescope of Purple Mountain Observatory in Delingha. The 3 mm cryogenically cooled 9 -- beam SIS (Superconductor Insulator Superconductor) receiver was used in double sideband mode, and the system temperatures ranged from 105 to 140 K during the observations. Using the fast Fourier transform spectrometer, the $^{12}$CO velocity resolution was 0.16 km s$^{-1}$, while the $^{13}$CO velocity resolution was 0.17 km s$^{-1}$. The rms was about 0.1 K at 1 min integration time, and the line detection limit about 0.15 K. The three CO lines were observed simultaneously. The HPBW was 60\arcsec{} at 110 GHz. The grid spacing of mapping observations was 30\arcsec, and the pointing accuracy was better than 10\arcsec. The average integration time of every point was one minute. The source was mapped using the on-the-fly mode of observation. The standard source W51 was checked roughly every two hours.

\section{RESULTS}
\subsection{Data reduction and exhibition}
Data reduction for H$_2$CO, H110$\alpha$, $^{12}$CO, and $^{13}$CO lines were done using CLASS and GREG, which are parts of GILDAS\footnote{%
  \tiny
GILDAS package was developed by IRAM (Institute de Radioastronomie Millim\'{e}trique). http://www.iram.fr/IRAMFR/GILDAS.}. To enhance comparison with the observation, we smoothed the $^{12}$CO and $^{13}$CO observations to 10\arcmin, and resampled them on the H$_2$CO observing grid. Sources observed are shown in Table B.1. The H110$\alpha$ data are reported in Table B.2. The parameters of the H$_2$CO and the 6 cm continuum are reported in Table B.3, while $^{12}$CO and $^{13}$CO data are reported in Table B.4. The H$_2$CO, H110$\alpha$, $^{12}$CO, and $^{13}$CO line spectra are shown in Fig.B.1. Line integral intensities of H$_2$CO, $^{12}$CO, and $^{13}$CO for four sources are shown in Figs.1, A.1, and A.3. Continuum and H110$\alpha$ distributions are shown in Fig.A.5.

\begin{figure}[t]
\vspace*{0.2mm}
\begin{center}
\includegraphics[width=8.2cm]{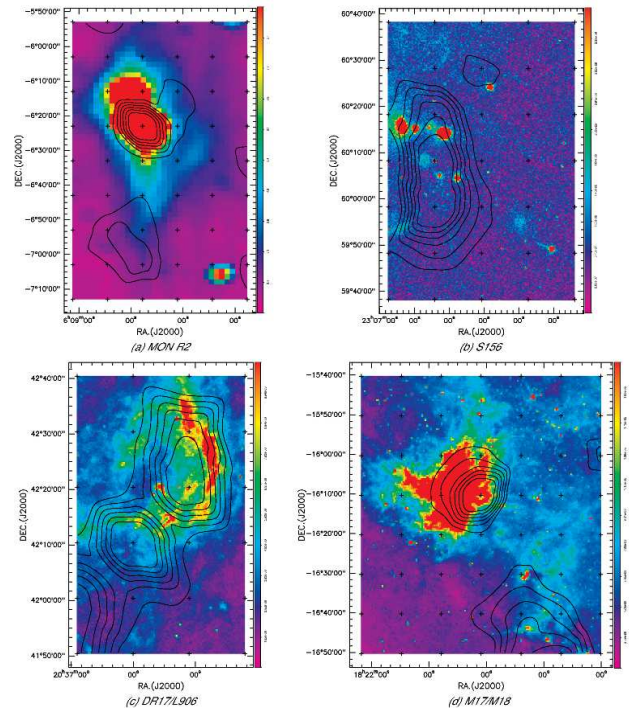}
\end{center}
\caption{Contours and color-scale maps of integrated area toward (a) MON R2, (b) S156, (c) DR17/L906, and (d) M17/M18. The contour and color-scale map respectively represent the integrated intensities of the H$_2$CO and the mid-infrared 8.28 $\mu$m MSX emission in S156, DR17/L906, and M17/M18 regions. For the MON R2 region , the color-scale map represents the IRAS 12 $\mu$m data. (a) For the MON R2 region: contour levels are -0.17 to -0.46 in steps of -0.06 K km s$^{-1}$. (b) For the S156 region: contour levels are -0.09 to -0.25 in steps of -0.03 K km s$^{-1}$. (c) For the DR17/L906 region: contour levels are -0.12 to -0.32 in steps of -0.04 K km s$^{-1}$. (d) For the M17/M18 region: contour levels are -0.42 to -1.11 in steps of -0.14 K km s$^{-1}$.}
\end{figure}

The H$_2$CO (1$_{10}$--1$_{11}$) apparent peak optical depth was determined using a simply standard radiative transfer result,
$\tau$$_{app}$ = -$\ln$[1 + T$_L$/(T$_c$ + 2.73 - T$_{ex}$)], where T$_L$ is the antenna temperature in K, T$_c$ is the continuum brightness temperature in K, and T$_{ex}$ is the excitation temperature of the 1$_{10}$--1$_{11}$ transition of H$_2$CO. The excitation temperature T$_{ex}$ values are in the range of 1.5 to 2.0 K (Heiles 1973, Vanden Bout et al. 1983, Young et al. 2004). Here we use a mean value of T$_{ex}$ = 1.7 K. The column density of H$_2$CO at rotational state 1$_{11}$ was obtained from the apparent peak optical depth using N(H$_{2}$CO) = 9.4 $\times$ 10$^{13}$$\cdot$$\tau$$_{app}$$\cdot$$\Delta$V (Pipenbrink \& Wendker 1988), where $\Delta$V is the FWHM in km s$^{-1}$. The difference between our assumed value of Tex and the one used in deriving the constant in the above expression (2.0 K) results in an offset of less than 0.1 dex in Fig.5. And the H$_2$ column density was obtained using the relation N(H$_2$CO)/N(H$_2$) = 1.25 $\times$ 10$^{-9}$ (Few \& Booth 1979, Scoville \& Solomon 1973, Evans et al. 1975). The local thermodynamic equilibrium (LTE) electron temperature T$_{e}$$^{*}$ was estimated following Brown et al. (1978) and Pipenbrink \& Wendker (1988). The optical depths of $^{13}$CO and the column densities of both $^{13}$CO and H$_2$ were estimated using Sato (1994) calculations, on the assumption that the cloud is in LTE.

\subsection{Description of sources}
MON R2. -- The size of MON R2 observed is about 60\arcmin{}$\times$90\arcmin{}. The H110$\alpha$ line was not detected. Maps of the integrated intensities of the H$_2$CO, $^{12}$CO, and $^{13}$CO line velocities range from 0 to 20 km s$^{-1}$. There is a velocity gradient of several km s$^{-1}$ for H$_2$CO, $^{12}$CO, and $^{13}$CO across the cloud. The spectrum of intensity peaks of H$_2$CO shows two velocity components at 7.6 and 10.52 km s$^{-1}$. The velocity component at 10.52 km s$^{-1}$ agrees with the intensity peaks of the $^{12}$CO and $^{13}$CO lines.

S156. -- About 50\arcmin{}$\times$70\arcmin{} of this region have been observed. We did not detect the H110$\alpha$ line. The line widths in the central part of the cloud are $\sim$ 3.8 km s$^{-1}$, more than twice the width normally measured in the dark clouds. This suggests that the cloud could be affected by the \ion{H}{ii} region. Maps of integrated intensities of H$_2$CO, $^{12}$CO, and $^{13}$CO line velocities range from -55 to -45 km s$^{-1}$. There is a velocity gradient of several km s$^{-1}$ for H$_2$CO, $^{12}$CO, and $^{13}$CO across the cloud. The velocity of the H$_2$CO intensity peak is -50.28 km s$^{-1}$, which agrees with those of the $^{12}$CO and $^{13}$CO emission lines.

\begin{figure}[t]
\vspace*{0.2mm}
\begin{center}
\includegraphics[width=7cm]{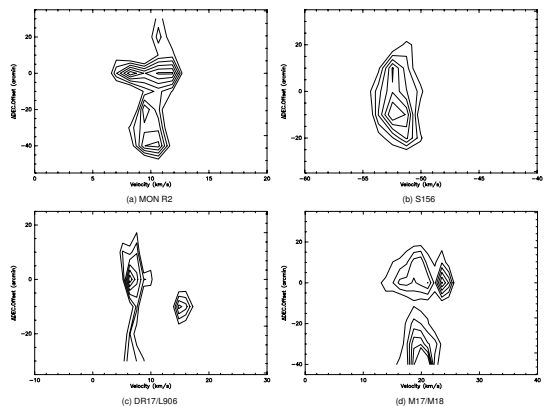}
\end{center}
\caption{Position-velocity diagrams of H$_2$CO for (a) MON R2 and (b) S156, regions along DEC. for $\Delta$$\alpha$ = 0, for (c) DR17/L906 region from offset (20, -30) to (-10, 20), and (d) M17/M18 region from offset (-30, -40) to (0, 30). For the MON R2, S156, and DR17/L906 regions: contour levels are -0.03 to -0.08 in steps of -0.01 K. For the M17/M18 region: contour levels are -0.03 to -0.18 in steps of -0.03 K.}
\end{figure}

DR17/L906. -- The size of our observed region is about 40\arcmin{}$\times$60\arcmin{}. The strong H110$\alpha$ emission was detected in the DR17 region with an average velocity of 11.4 km s$^{-1}$. Maps of the integrated intensities of the H$_2$CO, $^{12}$CO and $^{13}$CO line velocities range from -10 to 20 km s$^{-1}$. The H$_2$CO average spectrum shows two velocity components at 6.3 and 14.9 km s$^{-1}$. The velocity component at 6.3 km s$^{-1}$ of the H$_2$CO agrees with those of the DR17 \ion{H}{ii} region and the continuum emission. This is nearly half of the velocity of the H110$\alpha$ recombination line. The $^{12}$CO and $^{13}$CO average spectrum shows three velocity components at -1.6, 6.0, and 13.8 km s$^{-1}$. The $^{12}$CO and $^{13}$CO velocity components at 13.8 km s$^{-1}$ and the H$_2$CO velocity component at 14.9 km s$^{-1}$ are associated with the L906 region.

M17/M18. -- About 70\arcmin{}$\times$80\arcmin{} region have been observed. It covers the M17 region and north of M18. Nine H110$\alpha$ recombination lines were detected in the M17 region with an average velocity of 16.4 km s$^{-1}$. Maps of the integrated intensities of the H$_2$CO, $^{12}$CO and $^{13}$CO line velocities range from 10 to 50 km s$^{-1}$. The H$_2$CO average spectrum shows three velocity components at 18.3, 22.4, and 37.6 km s$^{-1}$. The $^{12}$CO and $^{13}$CO average spectrum shows four velocity components at 20.0, 28.2, 38.3, and 57.8 km s$^{-1}$. The H$_2$CO velocity component 18.3 km s$^{-1}$ agrees with the velocity component at 20.0 km s$^{-1}$ of the $^{12}$CO and $^{13}$CO, which also agrees with the \ion{H}{ii} region in M17. The H$_2$CO cloud distribution, line temperatures, and narrow line width comparing with values observed in the dark cloud shows that the cloud has been negatively influenced by the \ion{H}{ii} region for the velocity component at 22.4 km s$^{-1}$. The H$_2$CO velocity component at 37.6 km s$^{-1}$ agrees with the 38.3 km s$^{-1}$ value for the $^{12}$CO and $^{13}$CO lines.

\begin{figure}[t]
\vspace*{0.2mm}
\begin{center}
\includegraphics[width=5cm]{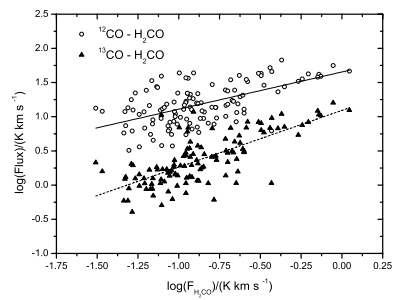}
\end{center}
\caption{Correlation between the H$_2$CO line flux and $^{12}$CO, $^{13}$CO line flux at corresponding points in MON R2, S156, DR17/L906, and M17/M18. The solid line is the linear fit for H$_2$CO and $^{12}$CO flux data, and the dashed line is the linear fit for H$_2$CO and $^{13}$CO.}
\end{figure}

\begin{figure}[t]
\vspace*{0.2mm}
\begin{center}
\includegraphics[width=5cm]{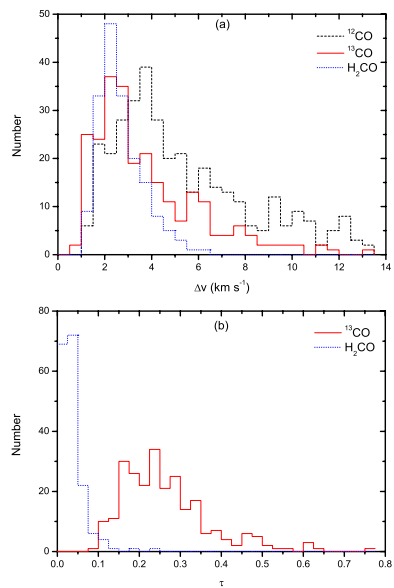}
\end{center}
\caption{(a) Histogram showing the measured widths for H$_2$CO, $^{12}$CO, and $^{13}$CO. (b) Histogram showing the histogram of the peak optical depths of H$_2$CO and $^{13}$CO distribution. Parts of H$_2$CO data are selected from Zhang et al. (2012).}
\end{figure}

\section{DISCUSSION}
We mapped four regions for the H$_2$CO, $^{12}$CO, and $^{13}$CO (Figs.1, A.1, and A.3). The maps show that the distribution of H$_2$CO is similar to $^{12}$CO and $^{13}$CO on a scale of 2 $\sim$ 10 pc. The position-velocity diagrams toward four regions (Figs.2, A.2, and A.4) also show that the H$_2$CO velocity agrees well with $^{12}$CO and $^{13}$CO. This suggests that H$_2$CO is directly related to the CO, and not merely located along the same line of sight. These indicate that in large-scale H$_2$CO can trace warm regions like $^{12}$CO and $^{13}$CO. The H$_2$CO is better correlated with the $^{13}$CO than with the $^{12}$CO especially in the Mon R2, S156, L906, and M17/M18 regions. The DR17 region shows a prominent feature that the $^{12}$CO and $^{13}$CO are not as strong as H$_2$CO. Toward four regions, the $^{12}$CO, $^{13}$CO, and H$_2$CO cloud distribution ranges gradually decrease with these three molecules. The H$_2$CO maps agree well with the IR maps toward the Mon R2 and DR17 regions. Towards the M17 region, the H$_2$CO intensity peak is at about 10$'$ ($\sim$4 pc) offset from the 8.28 $\mu$m MSX peak. The H$_2$CO absorption probably has no relation with the MSX emission in the S156 region.

The relation between H$_2$CO fluxes and those of $^{12}$CO and $^{13}$CO are shown in Fig.3. The best-fitted straight lines are
$\log$(F$_{^{12}CO}$) = (0.54 $\pm$ 0.07)$\log$(F$_{H_2CO}$) + (1.65 $\pm$ 0.07) (K km s$^{-1}$) and $\log$(F$_{^{13}CO}$) = (0.83 $\pm$ 0.07)$\log$(F$_{H_2CO}$) + (1.09 $\pm$ 0.07) (K km s$^{-1}$). The equations show that F$_{H_2CO}$ is linearly well correlated with  F$_{^{12}CO}$ and F$_{^{13}CO}$; the correlation coefficients are 0.58 and 0.73. The good relation between F$_{H_2CO}$ and F$_{^{13}CO}$ is an indication that the H$_2$CO and $^{13}$CO lines are located in similar environments.

Statistical histograms of the Gaussian fitting widths of H$_2$CO, $^{12}$CO, and $^{13}$CO (Fig.4 (a)) shows that the H$_2$CO line width range is from 1.2 to 8.5 km s$^{-1}$, and the average line width $<$$\Delta$V(H$_2$CO)$>$ is 2.5 km s$^{-1}$. All the H$_2$CO line widths observed exceed the thermal line widths 0.3 km s$^{-1}$. The thermal line widths for $^{12}$CO and $^{13}$CO is about 0.2 km s$^{-1}$, which is lower than all the $^{12}$CO and $^{13}$CO line widths observed. In addition, the widest hyperfine structure components of the H$_2$CO line are separated by about 0.8 km s$^{-1}$ (Young et al. 2004, Troscompt et al. 2009). This is less than our velocity resolution. The contribution of hyperfine structure broadening and thermal broadening to the measured H$_2$CO line widths is therefore likely to be small to moderate. Ninety percent of H$_2$CO line widths are distributed in the range of $\Delta$V(H$_2$CO) $<$ 4 km s$^{-1}$, and the distribution range of $\Delta$V(H$_2$CO) is less than for $^{12}$CO and $^{13}$CO. The main distributions and average line widths of H$_2$CO and $^{13}$CO are similar. There are no obvious correlations between the line widths of H$_2$CO, $^{12}$CO, and $^{13}$CO. The frequency distribution of the peak optical depths of H$_2$CO and $^{13}$CO spectra (Fig.4 (b)) shows that nearly all the H$_2$CO optical depths are lower than those of $^{13}$CO. The average optical depth of H$_2$CO is $<$$\tau$(H$_2$CO)$>$ $\sim$ 0.037, which can be compared to the value 0.055 quoted by Downes et al. (1980) for 262 galactic radio sources. The fairly low optical depth indicates that the H$_2$CO spectra is optically thin in almost all the regions observed.

The correlation between H$_2$CO and $^{13}$CO peak column density (Fig.5) shows that the H$_2$CO peak column density range is 1.1 $\times$ 10$^{12}$ -- 6.3 $\times$ 10$^{13}$ cm$^{-2}$, which is similar to the value range quoted by Federman et al. (1990) for the dense interstellar clouds. The column density corresponds to the MON R2, S156, L906, and M17 H$_2$CO intensity peak positions, and parts of M18 northern region are distributed in a box of the range N(H$_2$CO) $<$ 1.7 $\times$ 10$^{13} $cm$^{-2}$ and N($^{13}$CO) $>$ 4.7 $\times$ 10$^{15} $cm$^{-2}$, which shows a lack of H$_2$CO. With the exception of this box, the remaining data points show a strong correlation, and the best fit slope is N(H$_2$CO)/N($^{13}$CO) = (4.1 $\pm$ 0.2) $\times$ 10$^{-3}$. The N(H$_2$CO)/N($^{13}$CO) ratio against offset position towards four regions (Fig.6) shows that the N(H$_2$CO)/N($^{13}$CO) ratios increase from the center to the edge  regions of the molecular cloud where the N(H$_2$CO)/N($^{13}$CO) ratio increases about two to three times.

\section{CONCLUSIONS}
We observed and mapped large areas in four regions of MON R2, S156, DR17/L906, and M17/M18 using the H$_2$CO (1$_{10}$--1$_{11}$)  absorption, H110$\alpha$ recombination, 6 cm continuum, $^{12}$CO (1--0), and $^{13}$CO (1--0) emissions. The H$_2$CO distributions are similar to $^{12}$CO and $^{13}$CO distributions in four regions, with the $^{13}$CO distribution better correlated with the H$_2$CO distributions than the $^{12}$CO distribution. The H$_2$CO and $^{13}$CO tracers systematically provide consistent views of the dense regions, in which their maps have similar shapes, sizes, peak positions, and molecular spectra, presenting similar central velocities and line widths. Such good agreement indicates that the H$_2$CO absorption and the $^{13}$CO emission lines both arise in similar regions. The H$_2$CO and $^{13}$CO column density ratio N(H$_2$CO)/N($^{13}$CO) changes with different n(H$_2$) density regions in the molecular cloud. From the center to the edges of the molecular cloud, the N(H$_2$CO)/N($^{13}$CO) ratio jumps about two to three times.

\begin{acknowledgements}
We thank all the staff of Nanshan Observatory and Delingha of Purple Mountain Observatory for observations. And we thank Z. B. Jiang, Z. W. Chen, and J. Y. Li for providing the CO data of the M17/M18 region. This work was funded by The National Natural Science foundation of China under grant 10778703, and partly supported by the China Ministry of Science and Technology under State Key Development Program for Basic Research (2012CB821800) and The National Natural Science foundation of China under grant 10873025.
\end{acknowledgements} %

\begin{figure}[t]
\vspace*{0.2mm}
\begin{center}
\includegraphics[width=5cm]{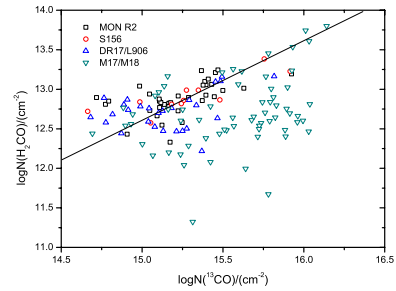}
\end{center}
\caption{Correlation between the H$_2$CO and $^{13}$CO peak column density. Straight line is the best-fit line for the source outside the range of N(H$_2$CO) $<$ 1.7 $\times$ 10$^{13} $cm$^{-2}$ and N($^{13}$CO) $>$ 4.7 $\times$ 10$^{15} $cm$^{-2}$.}
\end{figure}

\begin{figure}[t]
\vspace*{0.2mm}
\begin{center}
\includegraphics[width=5cm]{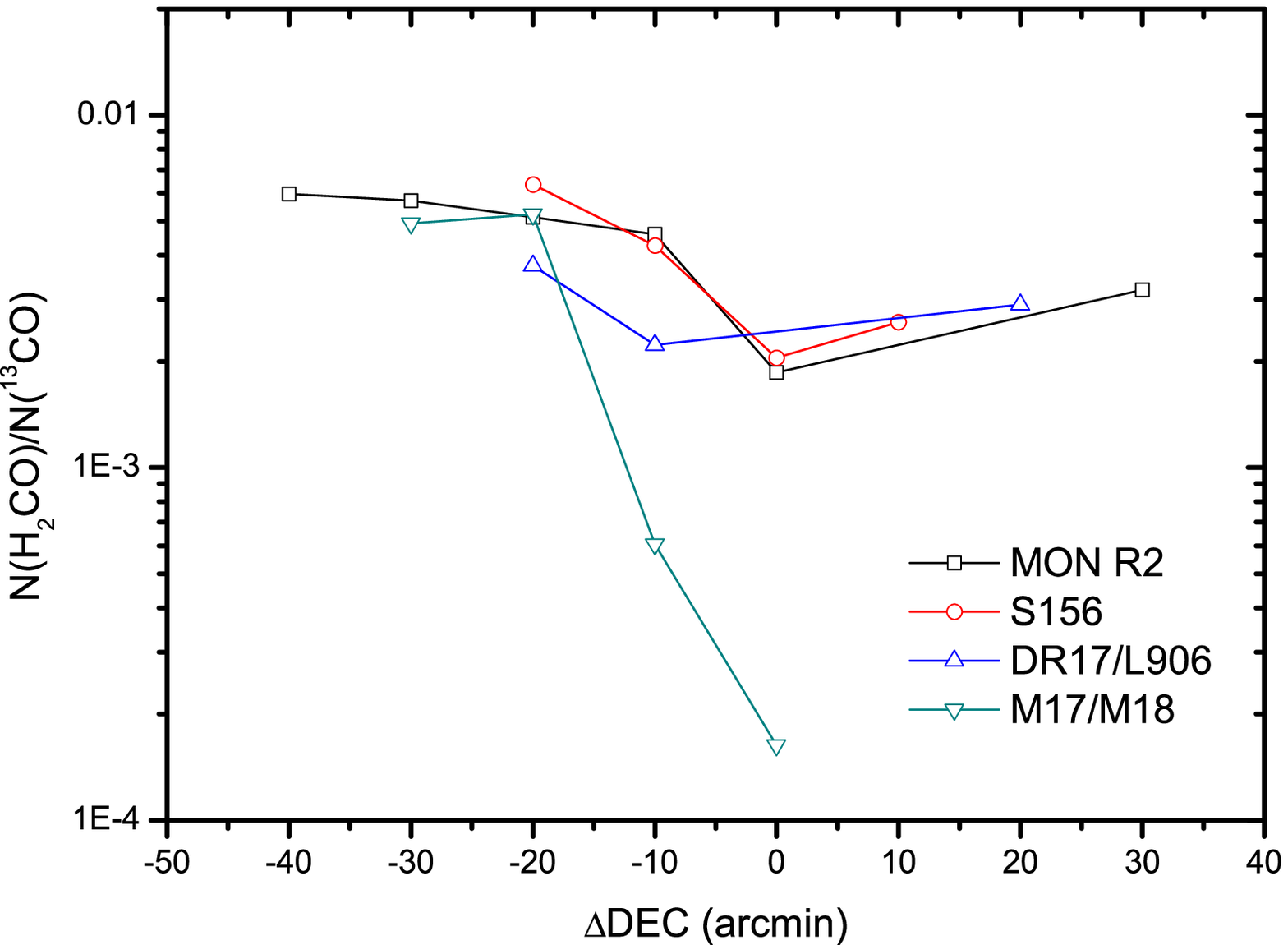}
\end{center}
\caption{Derived H$_2$CO and $^{13}$CO peak column density radio N(H$_2$CO)/N($^{13}$CO) against position along DEC. for $\Delta$$\alpha$ = 0 towards the MON R2, S156, and M17 regions. Towards DR17/L906 region $\Delta$$\alpha$ = 10.}
\end{figure}

\Online
\onecolumn
\begin{appendix} 

\section{The $^{12}$CO, $^{13}$CO and continuum maps}

\begin{figure}[htbp]
\vspace*{0.2mm}
\begin{center}
\includegraphics[width=10.2cm]{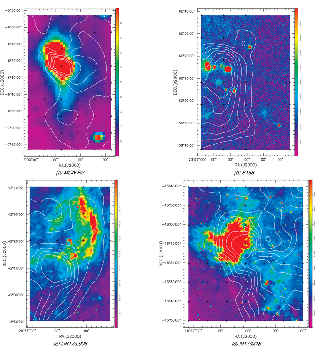}
\end{center}
\caption{The integrated intensities of the $^{12}$CO maps of integrated area toward (a) MON R2, (b) S156, (c) DR17/L906, and (d) M17/M18. The color-scale maps are the same as Fig.1. (a) For the MON R2 region: $^{12}$CO contour levels are from 12.80 to 34.14 in steps of 4.27 K km s$^{-1}$. (b) For the S156 region: $^{12}$CO contour levels are from 12.63 to 33.69 in steps of 4.21 K km s$^{-1}$. (c) For the DR17/L906 region: $^{12}$CO contour levels are from 12.72 to 39.26 in steps of 4.91 K km s$^{-1}$. (d) For the M17/M18 region: $^{12}$CO contour levels are from 29.34 to 78.24 in steps of 9.78 K km s$^{-1}$.}
\end{figure}

\begin{figure}[htbp]
\vspace*{0.2mm}
\begin{center}
\includegraphics[width=10.2cm]{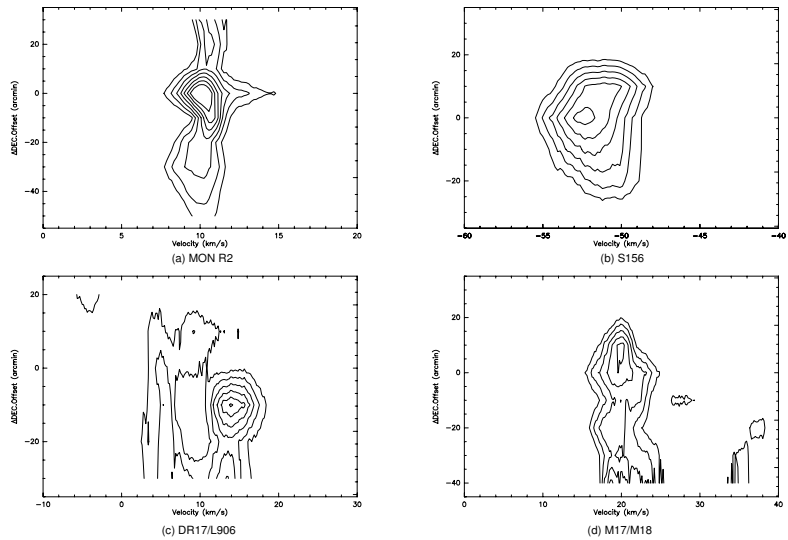}
\end{center}
\caption{Position-velocity diagrams of $^{12}$CO for (a) MON R2, (b) S156, (c) DR17/L906, and (d) M17/M18. The positions for $^{12}$CO are the same as Fig.2. For the MON R2, S156, and M17/M18 regions, contour levels are 3 to 8 in steps of 1 K. Contour levels are 1.3 to 6.3 in steps of 1 K for the DR17/L906 region.}
\end{figure}

\begin{figure}[htbp]
\vspace*{0.2mm}
\begin{center}
 \includegraphics[width=10.5cm]{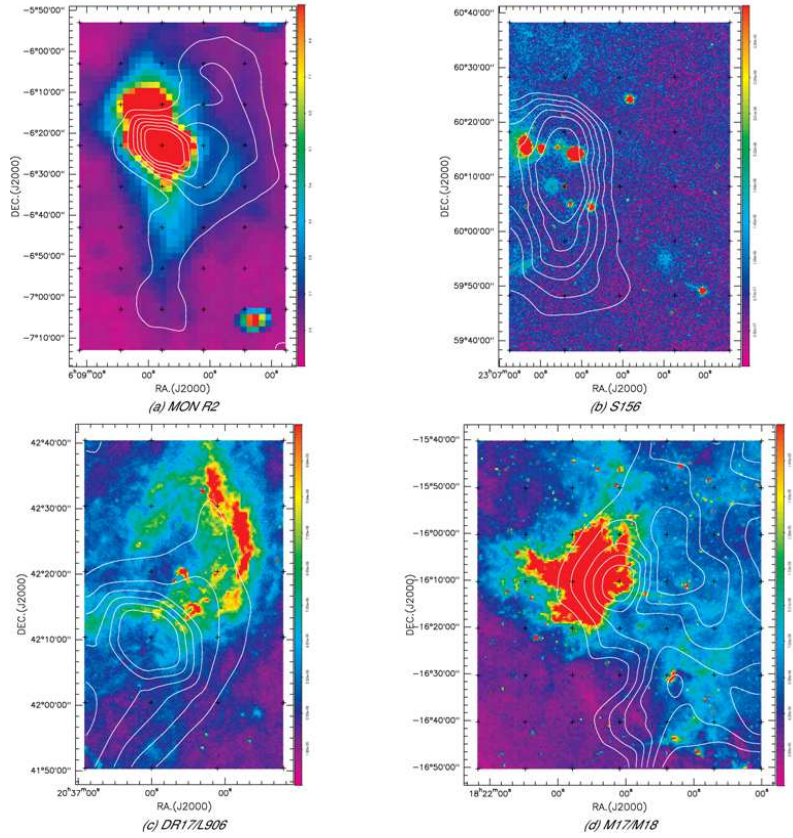}
\end{center}
\caption{The integrated intensities of the $^{13}$CO maps of integrated area toward (a) MON R2, (b) S156, (c) DR17/L906, and (d) M17/M18. The color-scale maps are the same as Fig.1. (a) For the MON R2 region: $^{13}$CO contour levels are from 2.66 to 7.10 in steps of 0.89 K km s$^{-1}$. (b) For the S156 region: $^{13}$CO contour levels are from 2.53 to 6.76 in steps of 0.84 K km s$^{-1}$. (c) For the DR17/L906 region: $^{13}$CO contour levels are from 2.64 to 7.05 in steps of 0.88 K km s$^{-1}$. (d) For the M17/M18 region: $^{13}$CO contour levels are from 7.83 to 20.89 in steps of 2.61 K km s$^{-1}$.}
\end{figure}

\begin{figure}[htbp]
\vspace*{0.2mm}
\begin{center}
\includegraphics[width=10.5cm]{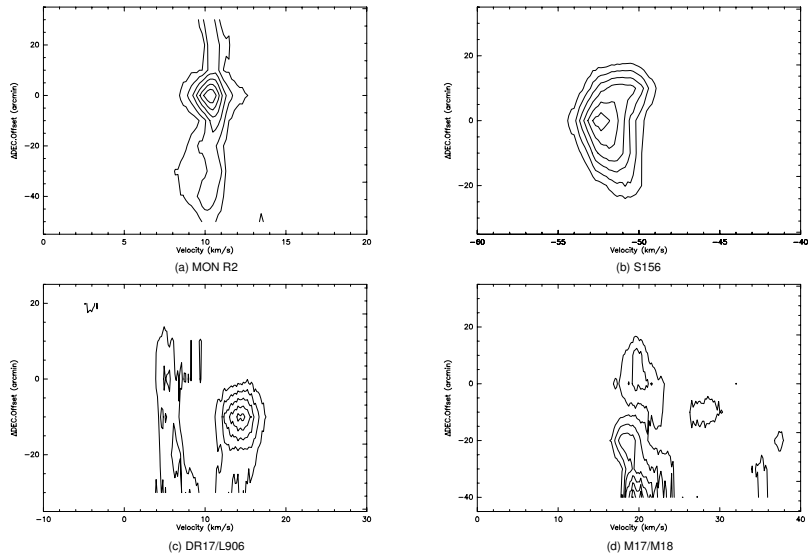}
\end{center}
\caption{Position-velocity diagrams of $^{13}$CO for (a) MON R2, (b) S156, (c) DR17/L906, and (d) M17/M18. The positions for $^{13}$CO are the same as in Fig.2. (a) For the MON R2 region: contour levels are 0.5 to 3 in steps of 0.5 K. (b) For the S156 region: contour levels are 0.6 to 2.1 in steps of 0.3 K. (c) For the DR17/L906 region: contour levels are 0.3 to 1.8 in steps of 0.3 K. (d) For the M17/M18 region: contour levels are 0.8 to 3.8 in steps of 0.5 K.}
\end{figure}

\begin{figure*}[htbp]
\centering
\includegraphics[width=12cm]{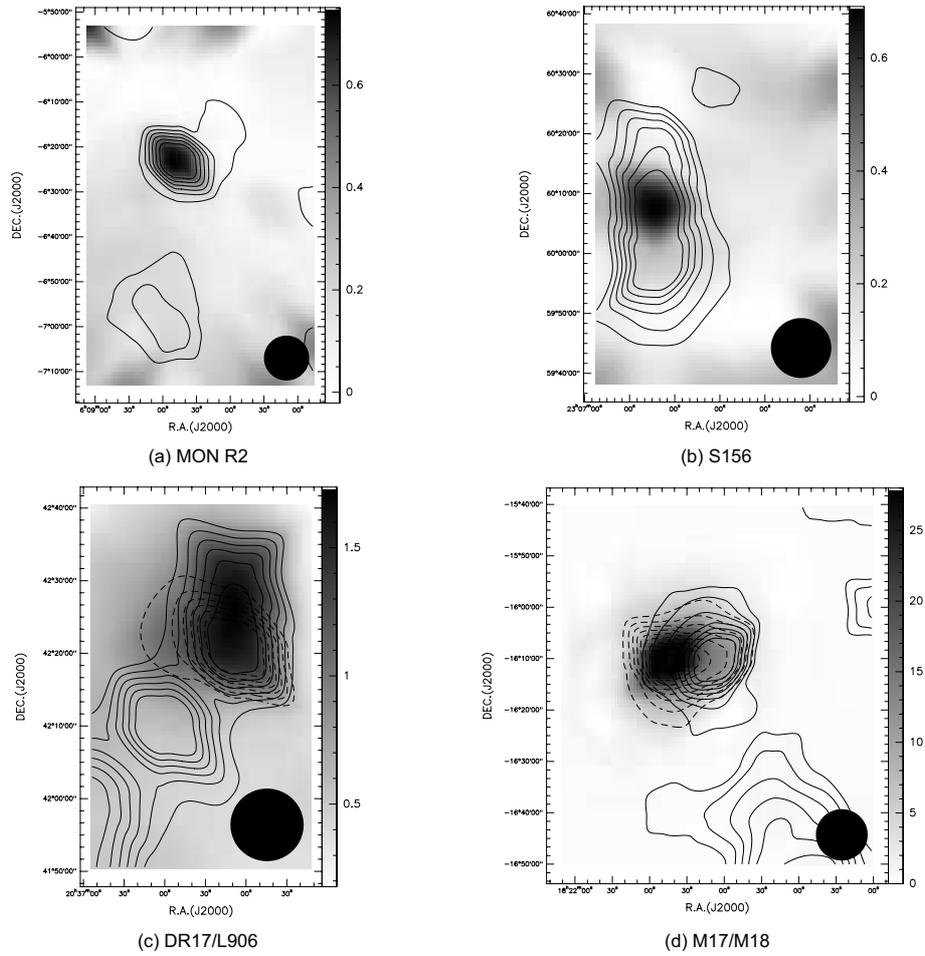}
\caption {Location of the radio continuum emission (gray scale) overlaid on the H$_2$CO observation lines (black contour lines) and H110$\alpha$ emission lines (dash contour lines) toward (a) MON R2, (b) S156, (c) DR17/L906, and (d) M17/M18. H$_2$CO contour levels are the same as Fig.1 for four regions. For the DR17/L906 region: H110$\alpha$ contour levels are 0.67, 0.89, 1.11, 1.33, 1.56, and 1.78 K km s$^{-1}$; for the M17/M18 region: H110$\alpha$ contour levels are 2.91, 4.36, 5.81, 7.26, 8.72, 10.17, and 11.62 K km s$^{-1}$. The gray bars are given in units of K.} \label{18}
\end{figure*}

\onecolumn
\section{Line parameters and spectra}

\begin{figure*}[htbp]
 \centering{
  \includegraphics[width=14cm,angle=0]{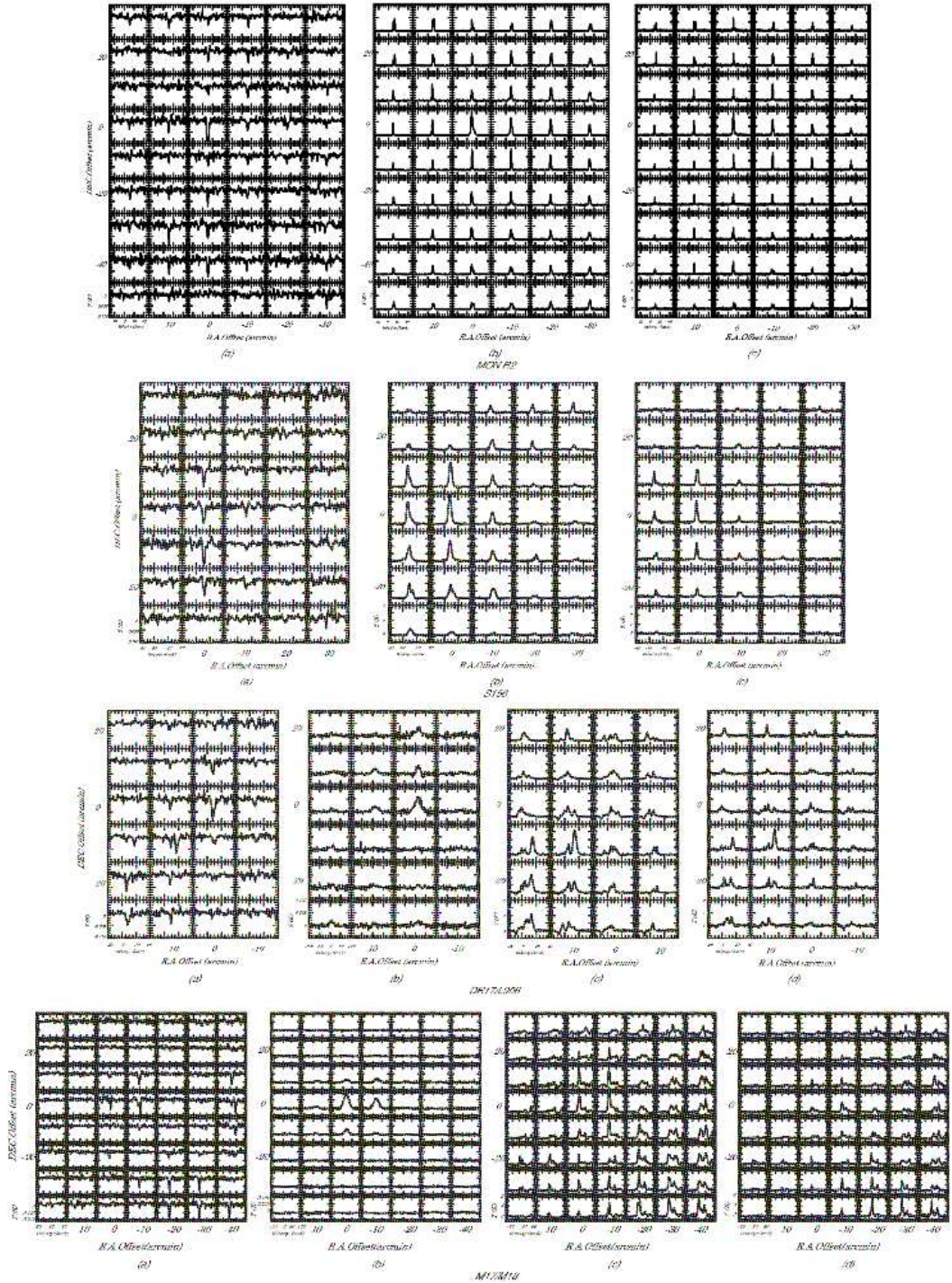}}
\caption{The spectra of (a) H$_2$CO, (b) $^{12}$CO, and (c) $^{13}$CO lines toward MON R2 and S156 regions. And the spectra of (a) H$_2$CO, (b) H110$\alpha$, (c) $^{12}$CO, and (d) $^{13}$CO lines toward DR17/L906 and M17/M18 regions.}
\label{fig:1}
\end{figure*}

\clearpage
 \onecolumn
  \begin{table}
   \caption{Source positions.}
    \label{table:1}
     \centering
\begin{tabular}
{c c c c c }
\hline\hline 
Sources & R.A.(J2000) & DEC.(J2000) & Size($\alpha$ $\times$ $\delta$) & References \\
& ($^{h}$ {} $^{m}$ {} $^{s}$) & (\degr {} \arcmin {} \arcsec) & (arcmin$^2$) & \\
\hline 
MON R2  & 06:07:46.60 &  -06:22:59.0 & 60\arcmin{}$\times$90\arcmin  & (1) \\
S156    & 23:05:24.80 &   60:08:14.0 & 50\arcmin{}$\times$70\arcmin  & (2) \\
DR17/L906 & 20:35:06.16 & 42:20:23.7 & 40\arcmin{}$\times$60\arcmin  & (3) \\
M17/M18     &  18:20:47.11 & -16:10:17.5 & 70\arcmin{}$\times$80\arcmin  & (4) \\
\hline 
 \end{tabular}
  \tablebib{(1) Herbst \& Racine (1976); (2) Hoglund \& Gordon (1973); (3) Schneider et al. (2006); (4) Downes et al. (1980).}
   \end{table}

\begin{table}
\caption{Parameters of the H110$\alpha$ recombination line.}
\label{table:1}
\centering
\begin{tabular}
{c c c c c c c c }
\hline\hline
Sources & Offset & Flux & Velocity & Width & T$_L$ & T$_C$ & T$_{e}$$^{*}$\\
& (arcmin) & (K km s$^{-1}$) & (km s$^{-1}$) & (km s$^{-1}$) & (K) & (K) & (K)\\
\hline
DR17/L906   &  0, 0    &  2.04(0.11)   & 12.7(0.8) & 31.5(1.9)  & 0.06 & 1.63 & 8154  \\
            &  0, 10   &  0.70(0.06)   & 14.7(0.9) & 18.2(1.6)  & 0.04 & 1.61 & 20320 \\
M17/M18     &  0, 0    &  14.40(0.20)  & 17.5(0.2) & 34.6(0.4)  & 0.39 & 31.1 & 19187 \\
            & -10, 0   &  9.91(0.25)   & 17.7(0.4) & 34.9(1.0)  & 0.27 & 13.9 & 13176 \\
            & -10, -10 &  1.24(0.08)   & 17.2(0.8) & 26.6(1.8)  & 0.04 & 0.95 & 7894  \\
            &  0, -10  &  3.89(0.10)   & 19.2(0.3) & 25.6(0.8)  & 0.14 & 6.58 & 15539 \\
            &  10, -10 &  1.35(0.14)   & 19.0(1.4) & 50.0(1.4)  & 0.03 & 1.34 & 9846  \\
            &  10, 0   &  1.92(0.10)   & 17.2(0.9) & 32.7(2.2)  & 0.06 & 2.36 & 11745 \\
            &  10, 10  &  1.17(0.09)   & 22.3(1.2) & 31.1(2.8)  & 0.04 & 1.81 & 14405 \\
            &  0, 10   &  2.72(0.13)   & 18.8(0.8) & 33.1(1.9)  & 0.08 & 5.96 & 19368 \\
            & -10, 10  &  2.72(0.13)   & 20.1(0.7) & 29.2(1.5)  & 0.09 & 4.99 & 16669 \\
\hline
\end{tabular}
\tablefoot{Parameters listed concern the simultaneously observed the H110$\alpha$ recombination line.}
  \end{table}

   \centering
   \begin{longtable}{ccccccccc}
    \caption{Parameters of the H$_2$CO (1$_{10}$--1$_{11}$) absorption line.} \\
     \hline\hline
  Sources & Offset & Flux & Velocity & Width & T$_L$ & T$_C$ \\
  & (arcmin) & (K km s$^{-1}$) & (km s$^{-1}$) & (km s$^{-1}$) & (K) & (K)\\
 \hline
  \endfirsthead
   \caption{continued.}\\
    \hline\hline
  Sources & Offset & Flux & Velocity & Width & T$_L$ & T$_C$ \\
  & (arcmin) & (K km s$^{-1}$) & (km s$^{-1}$) & (km s$^{-1}$) & (K) & (K) \\
 \hline
  \endhead
   \hline
    \endfoot
MON R2 &   0,    0 &  -0.30(0.04) & 7.6(0.2)  & 2.6(0.3) & -0.11   & 0.74     \\
       &           &  -0.24(0.04) & 10.5(0.1) & 1.9(0.4) & -0.12   &          \\
       & -10,    0 &  -0.21(0.03) & 10.7(0.3) & 3.8(0.8) & -0.05   &          \\
       & -10,  -10 &  -0.13(0.02) & 10.6(0.2) & 2.2(0.5) & -0.05   & 0.09     \\
       &   0,  -10 &  -0.15(0.03) & 9.7(0.3)  & 2.5(0.5) & -0.06   & 0.14     \\
       &  10,  -10 &  -0.08(0.02) & 9.3(0.1)  & 1.2(1.6) & -0.06   & 0.19     \\
       &  10,   0  &  -0.15(0.03) & 9.1(0.2)  & 1.9(0.6) & -0.07   & 0.16     \\
       &  10,   10 &  -0.05(0.02) & 9.9(0.2)  & 1.2(0.8) & -0.04   & 0.04     \\
\hline
 \end{longtable}
  \tablefoot{Parameters listed concern the simultaneously observed H$_2$CO absorption line. "N" indicates that the corresponding spectra could not be detected or the line peak intensity is less than 0.02 K.}

 \longtab{}{
   \begin{longtable}{cccccc|cccccc}
    \caption{Parameters of the $^{12}$CO (1--0) and $^{13}$CO (1--0) emission lines.}\\
     \hline\hline
      & & \multicolumn{4}{c}{$^{12}$CO} & \multicolumn{6}{c}{$^{13}$CO} \\
       \hline
  Sources & Offset & Flux & Velocity & Width & T$_L$ & Flux & Velocity & Width & T$_L$ \\
   & (arcmin) & (K km s$^{-1}$) & (km s$^{-1}$) & (km s$^{-1}$) & (K) & (K km s$^{-1}$) & (km s$^{-1}$) & (km s$^{-1}$) & (K)\\
 \hline
  \endfirsthead
   \caption{continued.}\\
    \hline\hline
     & & \multicolumn{4}{c}{$^{12}$CO}  &  \multicolumn{6}{c}{$^{13}$CO} \\
      \hline
  Sources & Offset & Flux & Velocity & Width & T$_L$ & Flux & Velocity & Width & T$_L$ \\
  & (arcmin) & (K km s$^{-1}$) & (km s$^{-1}$) & (km s$^{-1}$) & (K) & (K km s$^{-1}$) & (km s$^{-1}$) & (km s$^{-1}$) & (K)\\
 \hline
  \endhead
   \hline
    \endfoot
MON R2    &    0,    0 & 35.76(0.97) & 10.2  & 3.9  & 8.60 & 7.53(0.07)  & 10.3  & 2.2   & 3.26     \\
          &  -10,    0 & 24.62(0.12) & 10.7  & 3.9  & 6.01 & 4.73(0.05)  & 10.7  & 2.9   & 1.53     \\
          &  -10,  -10 & 21.28(0.11) & 10.7  & 2.5  & 8.00 & 4.29(0.04)  & 10.6  & 2.1   & 1.90     \\
          &    0,  -10 & 13.78(0.10) & 10.7  & 1.8  & 7.37 & 2.57(0.03)  & 10.6  & 1.4   & 1.71     \\
          &   10,  -10 & 6.60(0.07)  & 10.0  & 1.8  & 3.48 & 1.36(0.03)  & 10.0  & 1.3   & 0.99     \\
          &   10,   0  & 11.76(0.08) & 9.5   & 2.0  & 5.50 & 2.82(0.03)  & 9.5   & 1.6   & 1.62     \\
          &   10,   10 & 12.56(0.11) & 9.4   & 2.1  & 5.64 & 1.94(0.04)  & 9.3   & 1.5   & 1.23     \\
          &    0,   10 & 9.81(0.11)  & 10.3  & 2.5  & 3.65 & 1.92(0.06)  & 10.4  & 1.6   & 1.10     \\
\end{longtable}
 \tablefoot{Parameters listed concern the simultaneously observed $^{12}$CO and $^{13}$CO emission line. The $^{12}$CO and $^{13}$CO line velocities uncertainty together are about 0.2 km s$^{-1}$ and the line width uncertainty together are about 0.2 km s$^{-1}$.}
 }

\end{appendix}

\end{document}